\documentclass{PoS}
\usepackage{amsmath,amssymb,amsfonts}
\usepackage{slashed}
\usepackage{bbm}
\usepackage{cite}
\usepackage[utf8]{inputenc}
\usepackage[english]{babel}

\title{Lattice simulations of $G_2$-QCD at finite density}
\ShortTitle{$G_2$-QCD at finite density}
\author{\speaker{Bjoern H. Wellegehausen}\\%
        Justus-Liebig-University Giessen\\
        E-mail: \email{bjoern.wellegehausen@theo.physik.uni-giessen.de}}
\author{\speaker{Lorenz von Smekal}\\%
        Justus-Liebig-University Giessen and TU Darmstadt\\
        E-mail: \email{lorenz.smekal@physik.tu-darmstadt.de}}

\abstract{
$G_2$-QCD, in which the exceptional Lie group $G_2$ replaces the $SU(3)$ gauge group of QCD, does not suffer from a fermion sign problem. It can therefore be simulated also at comparatively low temperatures and high densities on the lattice, which hence allows to map out the phase diagram of this QCD-like theory.
We briefly review some of our previous results from such finite density simulations to then present further evidence for a first-order transition to what might be called $G_2$-nuclear matter. In order to isolate diquark condensation effects, we introduce simulations with Majorana fermions and diquark sources. 
This allows to disentangle states in the spectrum that are connected by charge conjugation. We discuss chiral symmetry in the presence of diquark sources and present first results from our ongoing large-scale simulations.
}

\FullConference{32nd International Symposium on Lattice Field Theory LATTICE 2014\\
		 Juny 23 - Juni 28, 2014\\
		 Columbia University, New York, USA}
\usepackage{etex}
\usepackage{graphicx}
\usepackage{bbm,bm}
\usepackage{mathtools}
\usepackage{color,array,tabularx}
\usepackage{stackrel}
\usepackage[small,bf]{caption}
\usepackage{pst-all}
\usepackage{psfrag}
\usepackage{pstricks,pst-grad,color,shadow}
\usepackage{tikz}
\usetikzlibrary{shapes,arrows,backgrounds}
\usepackage{soul}
\usepackage{placeins}
\usepackage{latexsym}
\usepackage{keyval}
\usepackage{simplewick}
\usepackage{epigraph}
\usepackage{amsthm}

\definecolor{darkgreen}{rgb}{0,0.73,0}

\newcommand{\erw}[1]{\left \langle #1 \right \rangle}

\newcommand{\id}{\mathbbm{1}}

\newcommand{\trnsp}{\mathsf{T}}

\DeclareMathOperator{\tr}{tr}

\newcommand{\ii}{\mathrm{i}}

\newcommand{\chargec}{\mathsf{C}}

\newcommand{\chk}{\textcolor{green}{\checkmark}}
\newcommand{\uchk}{\textcolor{red}{x}}

\graphicspath{{./}{./plots/}{./publPlots/}}

\begin{document}

\section{Introduction}

\noindent $G_2$-QCD is a QCD-like theory in which the gauge group $SU(3)$ of strong interactions is replaced by the exceptional Lie group $G_2$.
The theory is a gauge theory with fermionic baryons and 
fundamental quarks \cite{Holland:2003jy,Maas:2012wr} and it can be simulated without sign problem at finite density and temperature.
Unlike other QCD-like theories such as adjoint QCD or two color QCD, for example, its properties in the quenched  case are very similar to those of QCD \cite{Pepe:2006er,Greensite:2006sm,Danzer:2008bk,Wellegehausen:2010ai,Ilgenfritz:2012wg,Lacroix:2012pt}. Although the center of $G_2$ is trivial, it shows a first order
deconfinement transition which has quite interesting implications for the role of the center symmetry in QCD as reviewed in \cite{Maas:2012ts} for Lattice 2012.

In this contribution we briefly summarize our previous results \cite{Wellegehausen:2013cya} for the hadron spectrum in the vacuum and the phase structure at zero temperature as seen in the quark density from lattice simulations on rather  
small $8^3 \times 16$ lattices in Sections 2 and 3. The observed structures in the density over the chemical potential can thereby be related to a corresponding hierarchy of mass scales in the baryon spectrum. In particular, one observes 
thresholds in the baryon density at values of the chemical potential that 
correspond to the pseudo-Goldstone scalar diquark scale, 
an intermediate pseudo-scalar and vector diquark scale, and roughly the fermionic baryon scale set by the $G_2$-nucleons and delta baryons. In Section 4 we furthermore present results from our latest simulations in which we have accumulated evidence for a zero-temperature first-order phase transition to what might be called $G_2$-nuclear matter. 

The most important difference to QCD is the existence of diquarks in the hadronic spectrum. The lightest two diquark states are the pseudo-Goldstone bosons of chiral symmetry breaking,
and their quantum numbers differ only by charge conjugation. In order to investigate chiral symmetry breaking in detail we add diquark source terms to disentangle the charge
conjugation symmetry. Therefore we have to use Majorana fermions in the simulations. In Section 5 we review the formulation of $G_2$-QCD in terms of Majorana fermions and discuss the chiral properties
in the presence of diquark sources in more detail. First preliminary results of our simulations at finite temperature and density are shown in Sections 6 and 7. 

\section{Chiral symmetry and baryon number in $G_2$-QCD}\label{sg2props}

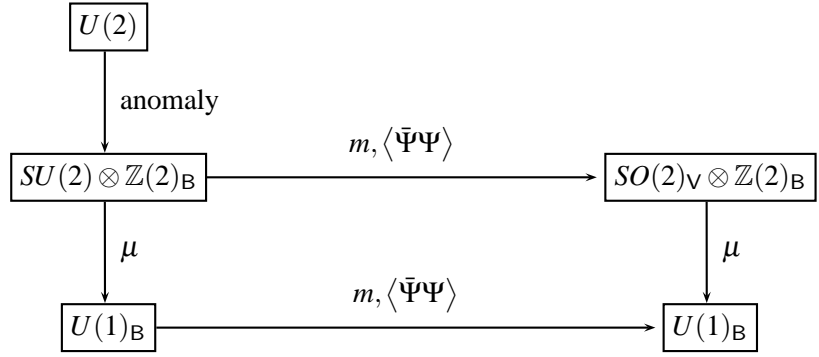
\begin{figure}[tb]
\psset{xunit=1cm,yunit=1cm,runit=1cm}
\begin{center}
\begin{pspicture}(0,0.5)(12,4.5)
\rput(2,4){\rnode{A}{
\psframebox{$U(2)$}}}
\rput(2,2){\rnode{B}{
\psframebox{$SU(2)
\otimes \mathbb{Z}(2)_\mathsf{B}$}}}
\rput(10,2){\rnode{C}{
\psframebox{$SO(2)_\mathsf{V}\otimes \mathbb{Z}(2)_\mathsf{B}$}}}
\rput(2,0){\rnode{D}{
\psframebox{$U(1)_\mathsf{B}$}}} 
\rput(10,0){\rnode{E}{
\psframebox{$U(1)_\mathsf{B}$}}}
\ncline{->}{A}{B}\naput{anomaly}
\ncline{->}{B}{C}\naput{$m, \erw{\bar{\Psi} \Psi}$}
\ncline{->}{C}{E}\naput{$\mu$}
\ncline{->}{D}{E}\naput{$m,\erw{\bar{\Psi} \Psi}$}
\ncline{->}{B}{D}\naput{$\mu$}
\end{pspicture}
\end{center}
\vskip5mm
\caption{Pattern of chiral symmetry breaking in $G_2$-QCD.}
\label{fig:chiralSymmG2QCD}
\end{figure}

\noindent The Euclidean action of $N_\text{f}=1$ flavour $G_2$-QCD with quark chemical potential $\mu$ for baryons reads
\begin{equation}
\begin{aligned}
S=&\int d^4 x \left \lbrace-\frac{1}{4} \tr F_{\mu\nu}F^{\mu\nu} +\bar{\Psi} \, D[A,m,\mu] \, \Psi\right \rbrace \quad \text{with} \\ 
D[A,m,\mu]=&\gamma^\mu (\partial_\mu-g A_\mu)-m +\gamma_0 \mu,
\label{eqn:actionQCD}
\end{aligned}
\end{equation}
where the gauge group is the exceptional Lie group $G_2$, and $\gamma_\mu=\gamma_\mu^\dagger$ are the Euclidean $\gamma$-matrices. The fundamental representations of $G_2$ are $7$-dimensional and 
$14$-dimensional, the latter coinciding with the adjoint representation. Since $G_2$ is a subgroup of $SO(7)$, all representations are real. 
The Dirac operator satisfies
\begin{equation}
D(\mu)^\dagger \,\gamma_5=\gamma_5\,D(-\mu^*) \quad \text{and} \quad D(\mu)^* \,T=T\,D(\mu^*)
\end{equation}
with $T=C\gamma_5$, $ T^*\, T=-\mathbbm{1}$, $T^\dagger=T^{-1}$ and charge conjugation matrix $C$. If such a unitary operator $T$ exists then the
eigenvalues of the Dirac operator come in complex conjugate pairs, all real
eigenvalues are doubly degenerate 
\cite{Kogut:2000ek,Hands:2000ei} and thus
\begin{equation}
\det D[\,A,m,\mu]\geq 0 \quad \text{for} \quad \mu \in \mathbbm{R}.
\end{equation}
$G_2$-QCD with a single Dirac flavour possesses an extended chiral symmetry \cite{Kogut:2000ek} compared to QCD. 
The action is invariant under the $SO(2)_\mathsf{V}$ vector transformations
and the usual axial transformations leading to a $U(2)$ symmetry group, in agreement with the results in \cite{Holland:2003jy}.
Following the same arguments as in QCD it is expected that 
the axial $U(1)$ is broken by the axial anomaly such 
that only a $SU(2)\times \mathbb{Z}(2)_\mathsf{B}$
chiral symmetry remains. In the presence of a non-vanishing Dirac mass term (or a non-vanishing chiral
condensate) the theory is no longer invariant under the axial
transformations. Therefore the non-anomalous chiral symmetry is
expected to be broken explicitly (or spontaneously) to its
maximal vector subgroup,
\begin{equation}
SU(2)\otimes
\mathbb{Z}(2)_\mathsf{B} \mapsto 
SO(2)_\mathsf{V}\otimes \mathbb{Z}(2)_\mathsf{B},
\end{equation}
The remaining chiral symmetry at finite baryon chemical potential is the same as in QCD,
\begin{equation}
SU(2)\otimes \mathbbm{Z}(2)_\mathsf{B} \mapsto 
U(1)_\mathsf{B}
\end{equation}
The final pattern of chiral symmetry breaking of $G_2$-QCD is shown in Figure \ref{fig:chiralSymmG2QCD}. If chiral symmetry is spontaneously broken, the axial chiral multiplet becomes
massless, according to the Goldstone theorem. The following operators generate the two Goldstone bosons:
\begin{equation}
\begin{aligned}
d(0^{++}) = \bar{\Psi}^\chargec \gamma_5 \Psi-\bar{\Psi}\gamma_5\Psi^\chargec \quad \text{and} \quad
d(0^{+-}) = \bar{\Psi}^\chargec\gamma_5\Psi+\bar{\Psi}\gamma_5\Psi^\chargec.
\end{aligned}
\end{equation}
They have quark number $n_\text{q} = 2$ and hence also carry a baryon number of $n_\text{B}=2/3$, if baryon number counts the difference of quarks and anti-quarks per $G_2$-nucleon as in QCD. The Goldstone bosons in $G_2$-QCD are scalar diquarks instead of
pseudoscalar mesons as in QCD. As long as we do not introduce diquark sources, $d(0^{++})$ and $d(0^{+-})$ have the same mass. In Section 5 below we also introduce the corresponding diquark source terms, however, in order to disentangle states with opposite charge quantum number and investigate chiral symmetry breaking more closely.

\section{Spectroscopy and the phase diagram at zero temperature}

\begin{figure}[b]
\begin{center}
\scalebox{0.9}{\input{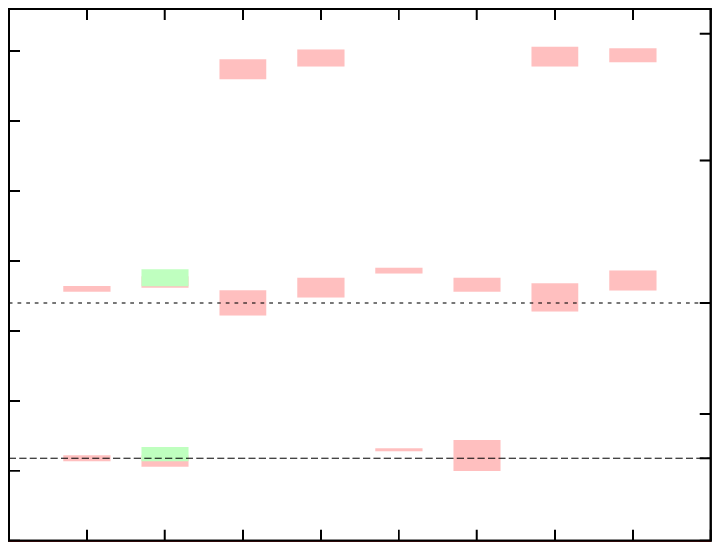}}\hskip05mm
\scalebox{0.9}{\input{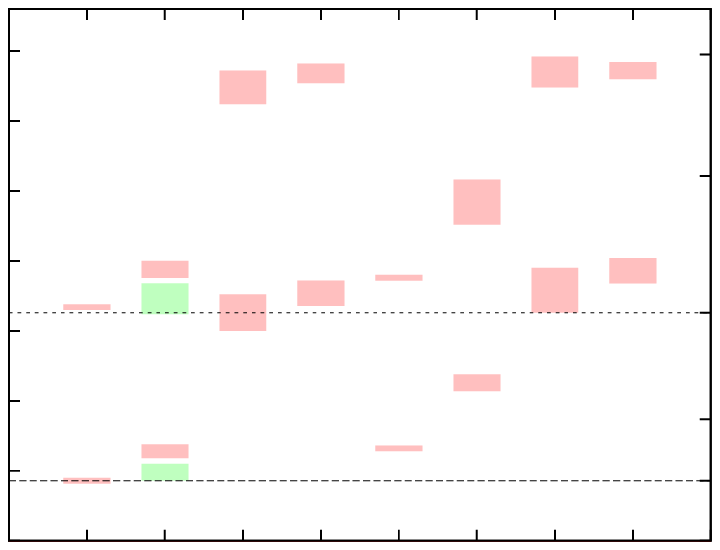}}
\end{center}
\caption{Mass spectrum for the heavy (left) and light (right) ensemble in $G_2$-QCD.}
\label{fig:mass}
\end{figure}

\begin{figure}[htb]
\begin{center}
\vskip10mm
\scalebox{0.9}{\input{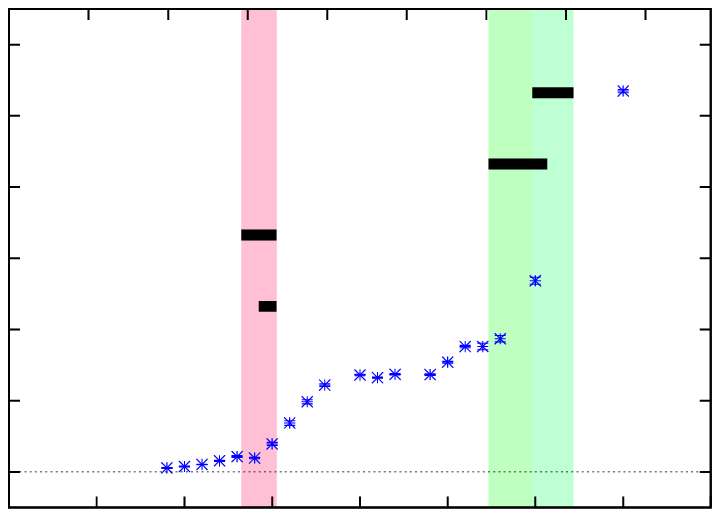}}\hskip05mm
\scalebox{0.9}{\input{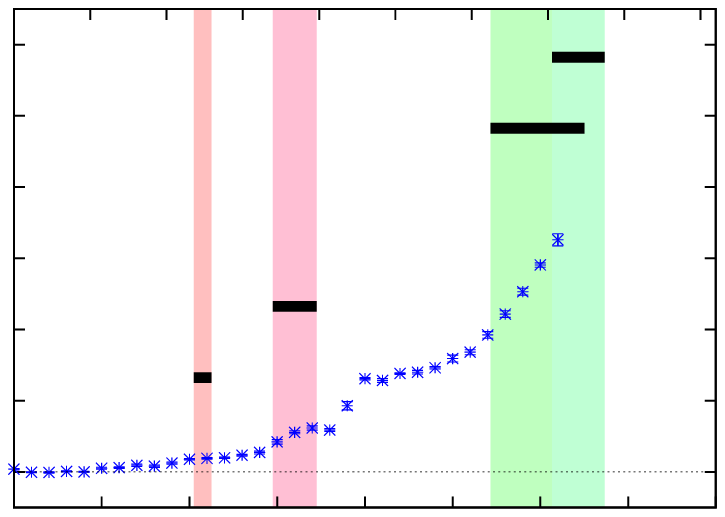}}
\end{center}
\caption{Quark number density for the heavy (left) and light (right) ensemble in $G_2$-QCD at finite density and zero temperature.}
\label{fig:qnd}
\end{figure}

\noindent The possible quark and gluon content of (colorless) bound states is determined by the tensor products of the appropiate representations of the gauge group $G_2$. 
Quarks in $G_2$ transform under the $7$-dimensional fundamental representation, gluons under the $14$-dimensional fundamental
(and at the same time adjoint) representation. An overview over possible bound states can be found in \cite{Wellegehausen:2013cya}.
We expect to find bound states for every integer quark number
$n_\text{q}$. Mesons have $n_\text{q}=0$, diquarks $n_\text{q}=2$, and nucleons $n_\text{q}=3$. 
In addition, there are more exotic bound states of gluons and quarks, for example a hybrid with $n_\text{q}=1$.
In the following we give an overview over the bound states considered here, where $u$ and $d$ denote flavours of Dirac fermions.
For the nucleons $N$ and the pion $\pi$ we make use of the partially quenched approximation in our one-flavour simulations.
Table \ref{tab:nb0} shows bound states that are also present in QCD while Table \ref{tab:nb1} shows the diquarks.
\begin{table}[htb]
\begin{minipage}{0.45\textwidth}
\begin{tabular}{|c|c|c|c|c|c|}
\hline Name & $\mathcal{O}$ & $T$ & J & P & C\\
\hline $\pi$ & $\bar{u} \gamma_5 d$ & SASS & 0 & - & + \\
\hline $\eta$ & $\bar{u} \gamma_5 u$ & SASS & 0 & - & + \\
\hline
\end{tabular}
\end{minipage}
\begin{minipage}{0.45\textwidth}
\begin{tabular}{|c|c|c|c|c|c|}
\hline Name & $\mathcal{O}$ & $T$ & J & P & C\\
\hline $N$ & $T^{abc}(\bar{u}_{a}^\chargec \gamma_5 d_{b}) u_{c}$ & SAAA & 1/2 & $\pm$ & $\pm$ \\
\hline $\Delta$ & $T^{abc} (\bar{u}_{a}^\chargec \gamma_\mu u_{b}) u_{c}$ & SSAS & 3/2 & $\pm$ & $\pm$ \\
\hline
\end{tabular}
\label{tab:nb3}
\end{minipage}
\caption{Bound states of $G_2$-QCD with $2$
flavours for baryon number $n_\text{B}=0$ (left) and baryon number $n_\text{B}=1$, i.e.\ quark number $n_\text{q} = 3$ (right). For details see text.}
\label{tab:nb0}
\end{table}
\begin{table}[htb]
\begin{center}
\begin{tabular}{|c|c|c|c|c|c|}
\hline Name & $\mathcal{O}$ & $T$ & J & P & C\\
\hline $d(0^{++})$ & $\bar{u}^\chargec \gamma_5 u + c.c.$ & SASS & 0 & + & + \\
\hline $d(0^{+-})$ & $\bar{u}^\chargec \gamma_5 u - c.c.$ & SASS & 0 & + & - \\
\hline $d(0^{-+})$ & $\bar{u}^\chargec u + c.c.$ & SASS & 0 & - & + \\
\hline $d(0^{--})$ & $\bar{u}^\chargec u - c.c.$ & SASS & 0 & - & - \\
\hline $d(1^{++})$ & $\bar{u}^\chargec \gamma_\mu d - \bar{d}^\chargec \gamma_\mu u + c.c.$ & SSSA & 1 & + & + \\
\hline $d(1^{+-})$ & $\bar{u}^\chargec \gamma_\mu d - \bar{d}^\chargec \gamma_\mu u - c.c.$ & SSSA & 1 & + & - \\
\hline $d(1^{-+})$ & $\bar{u}^\chargec \gamma_5 \gamma_\mu d - \bar{d}^\chargec \gamma_5 \gamma_\mu u + c.c.$ & SSSA & 1 & - & + \\
\hline $d(1^{--})$ & $\bar{u}^\chargec \gamma_5 \gamma_\mu d - \bar{d}^\chargec \gamma_5 \gamma_\mu u - c.c.$ & SSSA & 1 & - & - \\
\hline
\end{tabular}\\
\caption{Bound states with baryon number $n_\text{B}=2/3$, i.e.\ quark number $n_\text{q}=2$.}
\label{tab:nb1}
\end{center}
\end{table}
In all tables $\mathcal{O}$ is the interpolating operator used to extract the mass in simulations, the string $T$ represents the behaviour of the wave function under change of position, spin, colour and flavour (S stands for symmetric, A for anti-symmetric), and $J$, $P$, $C$ are
the spin, parity and charge conjugation quantum numbers. 
The difference between the $\eta$ and the diquark correlation function is only the disconnected contribution. 
Therefore, the diquark with positive parity has the same mass a the pion with negative parity, $m_{d(0^+)}=m_{\pi(0^-)}$.
In \cite{Wellegehausen:2013cya} it is shown that for every diquark there is a flavour non-singlet meson with the same mass but opposite parity. 
In the following we discuss two different ensembles with lattice parameters as listed in Table \ref{tab:ensembles}. 
A physical scale is set by the proton mass, $m_N=938$ MeV.
\begin{table}[htb]
\begin{tabular}{|c|c|c|c|c|c|c|c|c|}
\hline  Ensemble & $\beta$ & $\kappa$ & $m_{d(0^+)} a$ & $m_N a$  & $m_{d(0^+)}$ [MeV] & $a$ [fm] & $a^{-1}$ [MeV]& MC\\
\hline Heavy & $1.05$ & $0.147$ & $0.59(2)$ & $1.70(9)$ & $326$ & 0.357(33) & 552(50) & 7K   \\
\hline Light & $0.96$ & $0.159$ & $0.43(2)$ & $1.63(13)$ & $247$ & 0.343(45) & 575(75) & 5K \\
\hline
\end{tabular}\\
\caption{Parameters for the two different ensembles. All results are from $8^3\times 16$ lattices.}
\label{tab:ensembles}
\end{table}
The mass spectrum for both ensembles is shown in Figure \ref{fig:mass}. 
In the heavy ensemble the diquark masses and all parity even and odd states are almost degenerate.
In the light ensemble the diquark masses are no longer degenerate. 
We observe a significant mass splitting between parity even and odd states as well as between scalar and vector diquarks. 
Especially the Goldstone boson becomes the lightest state, with the $\eta$ also being somewhat heavier. 
For the nucleons we also observe different masses for parity even and odd states and the spin 1/2 and spin 3/2 representations. 
In particuclar, we find three clearly different scales in the light spectrum: a pseudo-Goldstone scale, an intermediate boson scale set by the remaining diquarks, and the nucleon scale set by the $N$ and $\Delta$ masses.
This mass hierarchy of the spectrum seems to be reflected in various structures
of the quark density at zero temperature which one might thus attribute to 
different bosonic and fermionic phases at finite density, see Figure \ref{fig:qnd}.  
With increasing chemical potential, the quark number density first remains consistent with zero until it very quickly rises to a very small but nonzero value.
When we compare the critical chemical potential $\mu_c$ for this onset transition to the mass of the lightest baryon $m_{d(0^+)}$, the pseudo-Goldstone $0^+$ diquark in our case, we find that numerically very good agreement with the expectation from the  Silver Blaze property, i.e.\ $\mu_c =m_{d(0^+)}/2 $. The ground state changes from the vacuum to a finite-density ground state only when the quark chemical potential reaches the mass of the lightest baryon divided by its quark number so that the corresponding excitation energy vanishes. For bosonic excitations one might expect Bose-Einstein-condensation in a continuous second-order quantum phase transition at $\mu_c =m_{d(0^+)}/2 $, without binding energy, and our data is certainly consistent with that. 
For larger values of the chemical potential plateaus develop where the quark number density remains almost constant. Especially in the light ensemble, the step towards the second plateau conicides with the mass of the 
heavier bosonic diquark states divided by their quark number. It appears that the two bosonic baryon mass scales are not sufficiently separated from each other to resolve these two distinct transitions in the heavier ensemble. 
 
At around $a\mu=0.6$ for the heavy ensemble and $a\mu=0.55$ for the light ensemble the quark number density starts increasing again and no further plateau is observed. 
This transition appears to coincide with the mass scale of nucleon and $\Delta$ divided by three.
In both ensembles the general pattern thus seems to be that the various transitions in the quark number density are related to the various baryon masses in units of  their quark number.  
While for bosonic baryons the density bends towards the zero-axis with plateaus
forming after each transition, at the scale of the fermionic baryons the quark number density is convex, as seen most clearly in the light ensemble with better separation of scales, and continues to further increase with increasing chemical potential until saturation sets in, eventually, when the lattice starts to get filled with the maximum number of quarks per site, i.e.\ at $a^3 n_q = 14$ here. This clearly is a lattice artifact beyond the range of any hadronic interpretation of the density and it is therefore not shown here again, see \cite{Wellegehausen:2013cya}.  

\begin{figure}[hb]
\begin{center}
\vskip10mm
\scalebox{0.9}{\input{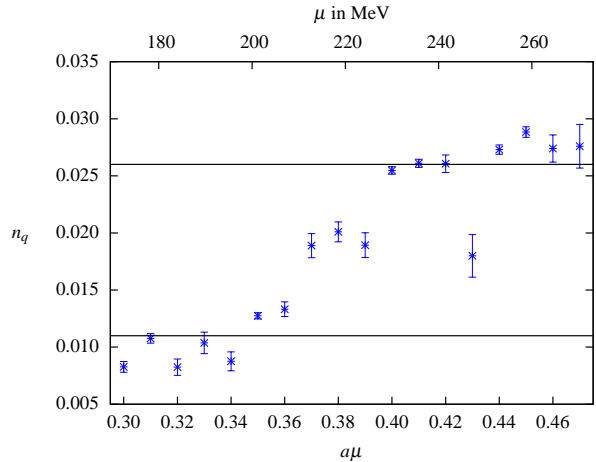}}
\end{center}
\caption{Quark number density for the light ensemble in the vicinity of the transition.}
\label{fig:qndDetail}
\end{figure}

\section{Evidence of a first order nuclear matter transition}

\begin{figure}[hb]
\begin{center}

\vspace*{.4cm}

\scalebox{0.62}{\input{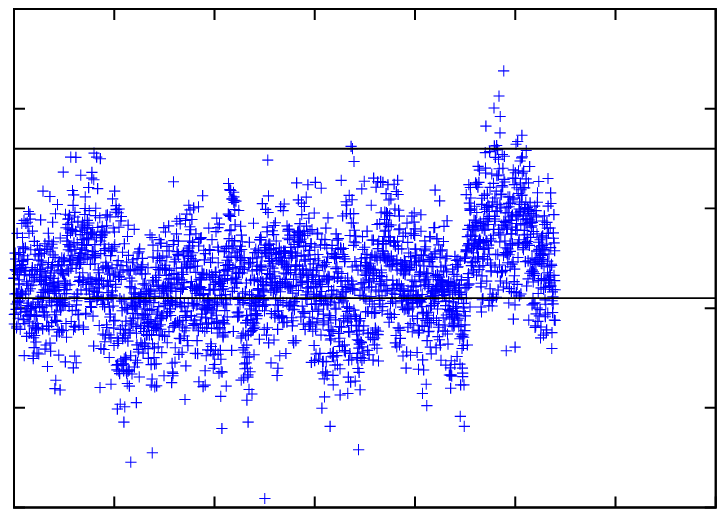}}
\scalebox{0.62}{\input{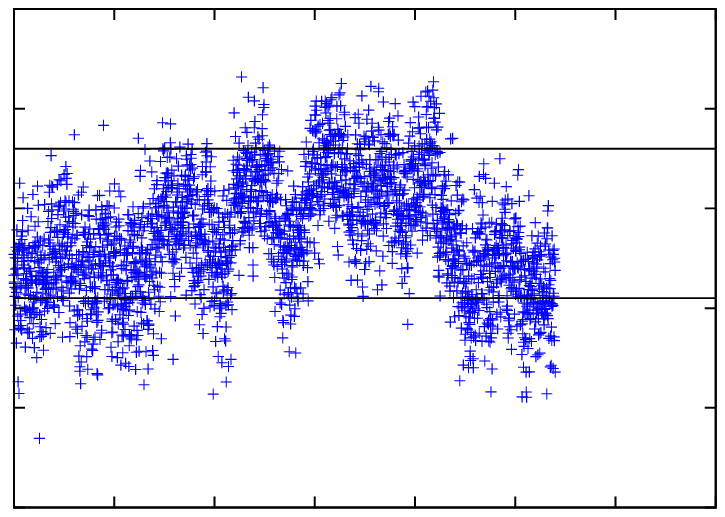}}
\scalebox{0.62}{\input{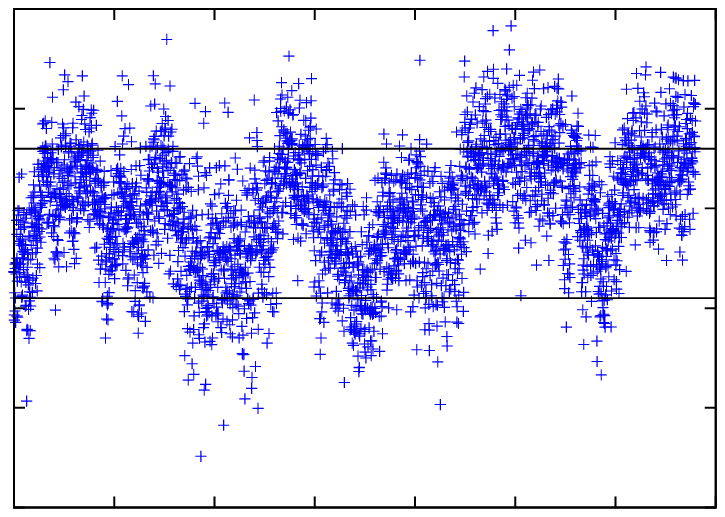}}
\scalebox{0.62}{\input{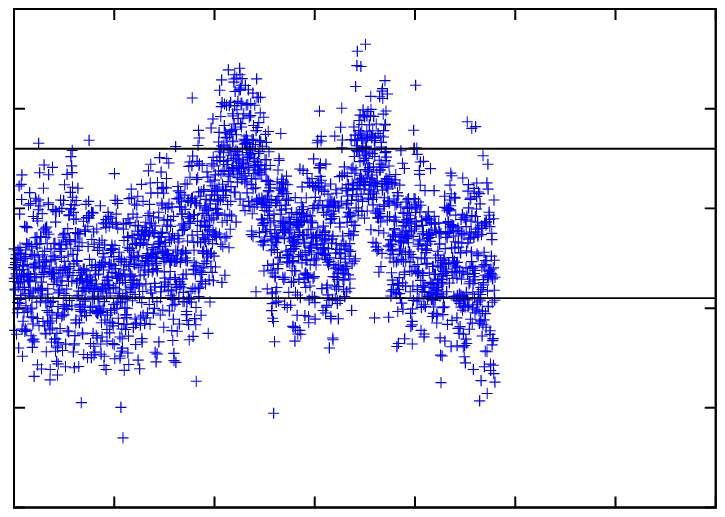}}
\scalebox{0.62}{\input{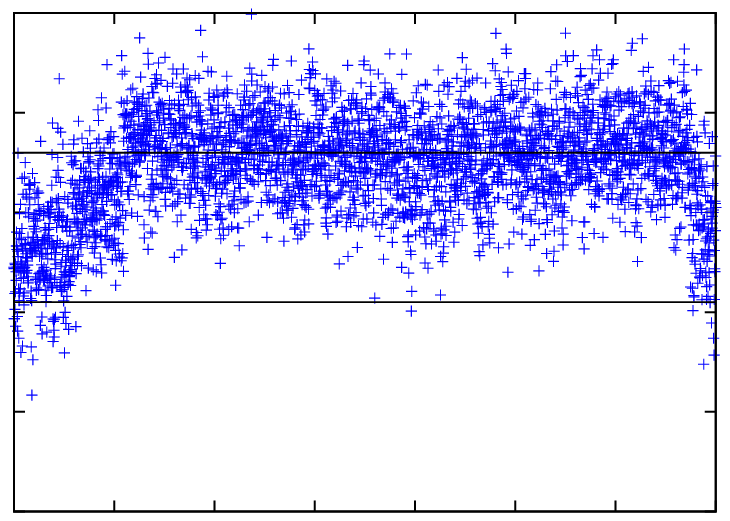}}
\scalebox{0.62}{\input{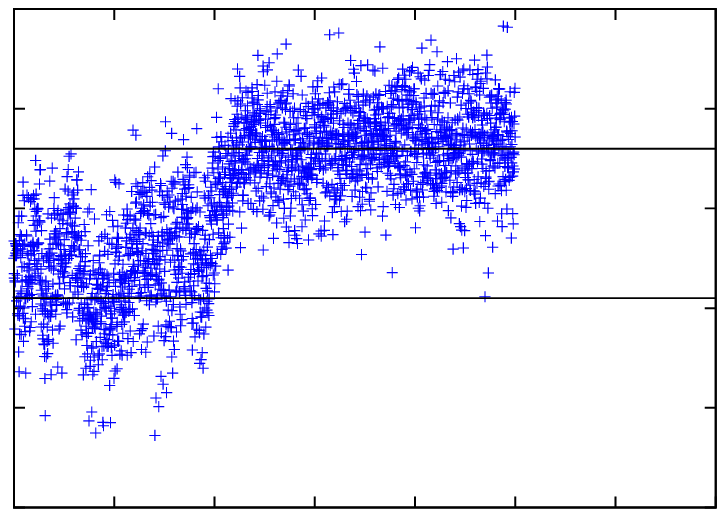}}
\caption{Quark number density as a function of Monte-Carlo time for different values of chemical potential. From left to right in the upper row $a\mu=0.36$, $a\mu=0.37$ and $a\mu=0.38$ and in the lower row $a\mu=0.39$, $a\mu=0.40$ and $a\mu=0.41$.}
\label{fig:qndMonteCarlo}

\end{center}
\end{figure}

In both ensembles we observe a strong transition at $a\mu\approx0.52$ (heavy ensemble) and $a\mu\approx0.38$ (light ensemble) that does not appear to correpond to any of our spectroscopic states. In Figure \ref{fig:qndDetail}
we show the vicinity of this transition in the light ensemble in more detail.
The quark number density rises between $a\mu=0.36$ and $a\mu=0.40$ from a lower value $n_q\approx 0.010$ to a higher value $n_q \approx 0.025$. In Figure
\ref{fig:qndMonteCarlo} we show the quark number density as a function of Monte-Carlo time and observe tunneling between these two states.
This might indicate that there is a first-order phase transition at $a \mu\approx 0.38$ in the phase diagram at zero temperature. Whether this phase transition is indeed the analogue of the liquid-gas transition of nuclear matter as expected in QCD remains to be shown by further simulations. 
If this is the case, then either the binding energy per nucleon is comparatively large or the masses of nucleon and  $\Delta$ change with density in the regime of the finite bosonic baryon density in the ground state before this transition which is not possible in QCD.

\section{Simulations of $G2$-QCD with Majorana fermions}

\noindent
In the present section we discuss the introduction of diquark sources in $G_2$-QCD for two reasons:
First, on larger lattices and especially for values of the chemical potential in the vicinity of the first order transition the simulations become more and more expensive. An obvious reason for this might be the presence of very light
diquark excitations in the simulations in this region of the phase diagram. With the introduction of diquark source terms simulations should become more feasible. Secondly, we would eventually like to resolve the complete diquark spectrum
in order to investigate chiral symmetry breaking at finite density. Any diquark source necessarily consists of an operator with quark number $n_\text{q}=2$  and therefore contains a pair of charge conjugated Dirac spinors $\Psi$ for the quarks. In order to  integrate over the fermion fields in the path integral, however, one needs a bilinear expression in $\bar\Psi$ and $\Psi $. Unlike two-color QCD, however, for a single flavor of Dirac fermions it is then necessary to introduce corresponding Majorana fermions in the simulations. Since the gauge fields in $G_2$ satisfy $A_\mu^\trnsp=-A_\mu=-A_\mu^a T_a$ it is possible to rewrite the matter part of the action in \eqref{eqn:actionQCD} as a sum over $2$ for $\mu=0$ at first uncoupled Majorana spinors $\lambda=(\chi\,,\eta)$, 
\begin{equation}
\begin{aligned}
S[\Psi,A]=&\int d^4x\,\bar{\Psi}\left(\gamma^\mu
(\partial_\mu-g A_\mu)-m \right)\Psi
=\int d^4x\,\bar{\lambda}\left(\gamma^\mu (\partial_\mu-g A_\mu)-m\right)\lambda
\label{eqn:actionQCDMaj}
\end{aligned}
\end{equation}
Here $\lambda$ obeys the Majorana condition $
\lambda^\chargec=C \bar{\lambda}^\trnsp=\lambda$, $
\bar{\lambda}^\chargec=-\lambda^\trnsp C^{-1}=\bar{\lambda}$,
and it is related to the Dirac spinor as
$\Psi=\chi+\ii\, \eta$, $\bar{\Psi}=\bar{\chi}-\ii\bar{\eta}$, $\Psi^\chargec=\chi-\ii\, \eta$ and $\bar{\Psi}^\chargec=\bar{\chi}+\ii\,
\bar{\eta}$.
The (baryon) chemical potential $\mu$ is an off-diagonal term in Majorana flavour space such that
\begin{equation}
\begin{aligned}
\mathcal{L}=&\bar{\Psi} \,D(m,\mu)\,\Psi=\bar{\lambda}\,M(m,\mu)\, \lambda \quad \text{with}\\
D(m,\mu)=&\slashed{D}-m + \gamma_0\, \mu \quad \text{and} \quad M(m,\mu)=(\slashed{D}-m) \sigma_0 -\mu \gamma_0 \sigma_2
\label{eqn:lagrangeQCDchem}
\end{aligned}
\end{equation}
where $\sigma_0=\mathbbm{1}$ and the Pauli matrix $\sigma_2$ act on the 2 Majorana flavours $\chi$ and $\eta$ which are now coupled to each other by the offdiagonal $\mu \sigma_2$-term.
Integration in the path integral over the Majorana fermions leads to the Pfaffian instead of the fermion determinant and we can proove that the Pfaffian is positive as expected.

\begin{figure}[htb]
\begin{center}
\begin{minipage}{0.60\textwidth}
\begin{pspicture}(1,-4)(15,4)
\rput(2,2){\rnode{B}{
\psframebox{$SU(2)_{\mathsf{L}=\mathsf{R}^*}$}}}
\rput(5,0.8){\rnode{C0}{
\psframebox{\textcolor{red}{$U(1)_\mathsf{B}$}}}}
\rput(8,2){\rnode{C1}{
\psframebox{\textcolor{blue}{$U(1)_\mathsf{B}$}}}}
\rput(2,-2){\rnode{D}{
\psframebox{\textcolor{red}{$U(1)_\mathsf{B}$}}}}
\rput(5,-1){\rnode{E}{
\psframebox{\textcolor{red}{$U(1)_\mathsf{B}$}}}}
\rput(8,-2){\rnode{F}{
\psframebox{-}}}
\ncline{->}{B}{C0}\nbput{$m$}
\ncline{->}{B}{C1}\naput{$\mathrm{lin}(m,\tilde{J}_1,\tilde{J}_2)$}
\ncline{->}{C0}{C1}\nbput{$\mathrm{lin}(\tilde{J}_1,\tilde{J}_2)$}
\ncline{->}{C0}{E}\naput{$\mu$}
\ncline{->}{C1}{F}\naput{$\mu$}
\ncline{->}{D}{E}\naput{$m$}
\ncline{->}{B}{D}\naput{$\mu$}
\ncline{->}{D}{F}\nbput{$\mathrm{lin}(m,\tilde{J}_1,\tilde{J}_2)$}
\ncline{->}{E}{F}\naput{$\mathrm{lin}(\tilde{J}_1,\tilde{J}_2)$}
\end{pspicture}
\end{minipage}
\begin{minipage}{0.25\textwidth}
\scalebox{0.35}{\includegraphics{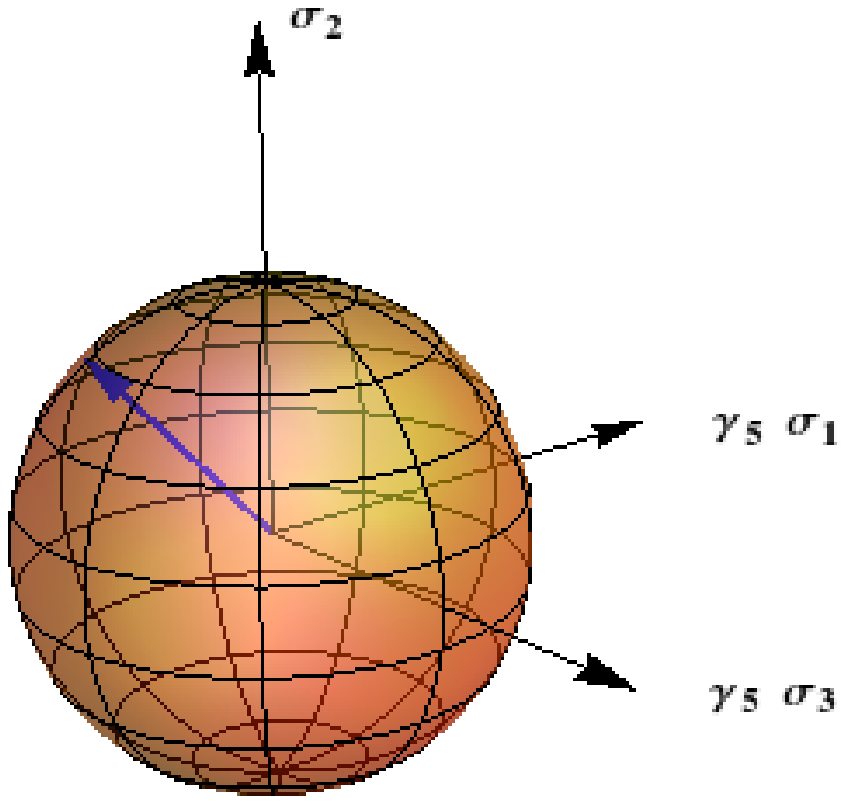}}\\
\vskip-18mm
\scalebox{0.35}{\includegraphics{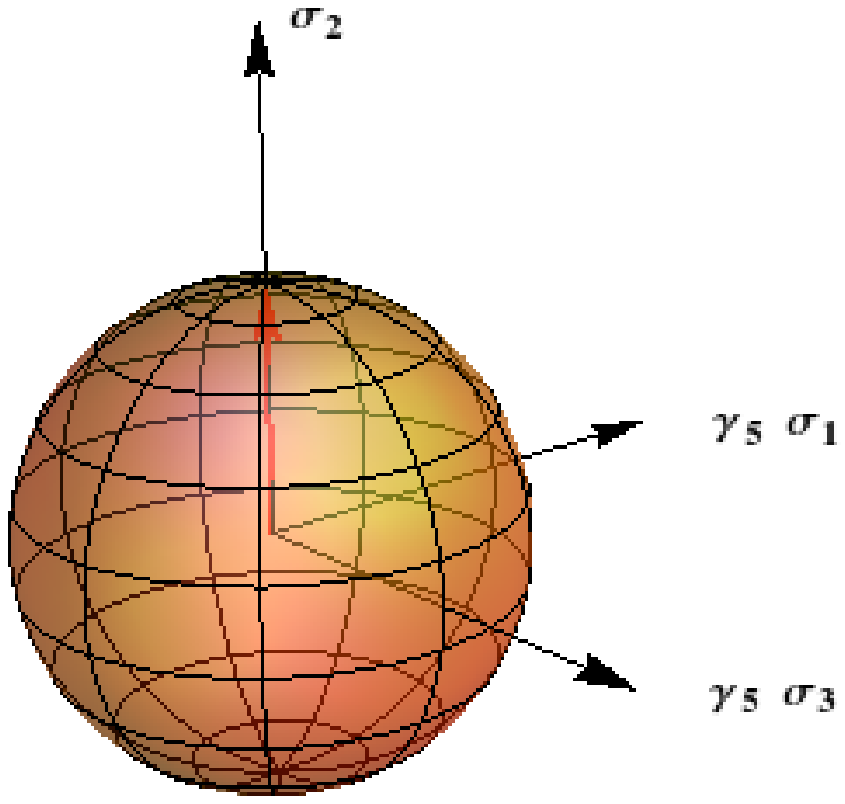}}
\end{minipage}
\end{center}
\caption{Chiral symmetry breaking in $G_2$-QCD in the presence of diquark sources.}
\label{fig:chiralSymmDiquark}
\end{figure}

It is possible to introduce two different diquark sources in $G_2$-QCD, one for the scalar ($\mathcal{L}_{\gamma_5} $) and one for the pseudoscalar diquarks ($ \mathcal{L}_\id$), 
\begin{equation}
\mathcal{L}_{\gamma_5}(\tilde{J})=\frac{1}{2}\left(\tilde{J} \bar{\Psi}^\chargec\gamma_5 \Psi-\tilde{J}^*\bar{\Psi}\gamma_5\Psi^\chargec\right)
\,,\quad 
 \mathcal{L}_\id(J)=\frac{1}{2}\left(J \bar{\Psi}^\chargec \Psi+J^*\bar{\Psi}\Psi^\chargec\right).
\end{equation}
In the Majorana decomposition these terms read ($J=J_1+\ii J_2$, $\tilde{J}=\tilde{J}_1+\ii \tilde{J}_2$)
\begin{equation}
\begin{aligned}
\mathcal{L}_{\gamma_5}=&\ii\, \begin{pmatrix} \bar{\chi} \\ \bar{\eta}\end{pmatrix}
\begin{pmatrix} J_2 & J_1 \\ J_1 & -J_2 \end{pmatrix} \,\gamma_5\, \begin{pmatrix} \chi \\\eta\end{pmatrix}=\ii \,\bar{\lambda} \left(\tilde{J_1} \sigma_1+J_2 \sigma_3 \right) \gamma_5\, \lambda
\,,\\
 \mathcal{L}_\id=&\begin{pmatrix} \bar{\chi} \\ \bar{\eta}\end{pmatrix}
\begin{pmatrix} J_1 & -J_2 \\ -J_2 & -J_1 \end{pmatrix} \begin{pmatrix} \chi \\\eta\end{pmatrix}=\bar{\lambda} \left(J_1 \sigma_3-J_2 \sigma_1 \right) \lambda\, . 
\end{aligned}
\end{equation}
The Lagrange density for the matter part of the theory is then given by
\begin{equation}
 \mathcal{L}=\bar{\lambda}\left[(\slashed{D}-m - m_5\gamma_5 )\sigma_0  -\mu \gamma_0 \sigma_2+ (\ii \tilde{J}_1  \gamma_5-J_2) \sigma_1+(J_1+\ii \tilde{J}_2 \gamma_5) \sigma_3\right]\lambda.
\end{equation}
Similar to the case without diquark sources one can show that the Pfaffian is real if $J_2=\tilde{J_2}=0$, but it is not necessarily positive any longer. Nevertheless, we expect that for small values of the chemical potential
the sign problem is not present and this expectation is confirmed in our simulations.
The first derivatives of the partition function with respect to $J$, $\tilde{J}$ and $m$ define the chiral and diquark condensates,
\begin{equation}
\begin{aligned}
 \Sigma=&\frac{1}{V} \frac{\partial \ln(Z(m,J,\tilde{J}))}{\partial m}=\erw{\bar{\Psi} \Psi}=\erw{\bar{\chi} \chi+\bar{\eta}\eta},\\
 \Sigma_1=&\frac{1}{V} \frac{\partial \ln(Z(m,J,\tilde{J}))}{\partial J}=\erw{\bar{\Psi}^\chargec \Psi + c.c.}=\erw{\bar{\chi} \chi-\bar{\eta}\eta},\\
 \Sigma_5=&\frac{1}{V} \frac{\partial \ln(Z(m,J,\tilde{J}))}{\partial \tilde{J}}=\erw{\bar{\Psi}^\chargec \gamma_5 \Psi + c.c.}=\ii \erw{\bar{\chi} \gamma_5 \eta+\bar{\eta} \gamma_5 \chi},
 \end{aligned}
\end{equation}
that we investigate in the following.
For a single Dirac flavour the chiral symmetry is $SU(2)_{\mathsf{L}=\mathsf{R}^*}$.
The generators for the symmetry transformations are given by Pauli matrices
\begin{equation}
 T_\mathsf{V}=\id \otimes \sigma_2\,,\quad T_\mathsf{A}=\gamma_5 \otimes \left\{\sigma_1 ,\sigma_3\right\},
\end{equation}
and the chiral transformations read
\begin{equation}
\begin{aligned}
 O_\mathsf{A,1} \, \lambda=&e^{\ii \,\alpha \, \gamma_5 \,\sigma_1}\,\lambda \quad \text{and}\quad \bar{\lambda} \rightarrow \bar{\lambda} \, e^{\ii \,\alpha \, \gamma_5 \,\sigma_1}\\
 O_\mathsf{A,3} \, \lambda=&e^{\ii \,\alpha \, \gamma_5 \,\sigma_3}\,\lambda \quad \text{and}\quad \bar{\lambda} \rightarrow \bar{\lambda} \, e^{\ii \,\alpha \, \gamma_5 \,\sigma_3}\\
 O_\mathsf{V,2} \, \lambda=&e^{\ii \,\alpha \, \sigma_2}\,\lambda \quad \text{and}\quad \bar{\lambda} \rightarrow \bar{\lambda} \, e^{-\ii \,\alpha \, \sigma_2}.
 \end{aligned}
\end{equation}
Possible bilinear bound states for a single Dirac flavour are
\begin{equation}
 \begin{aligned}
  d(0^{+-})=&\bar{\lambda} \gamma_5 \sigma_1 \lambda=\bar{\chi} \gamma_5 \eta\,,\quad d(0^{++})=\bar{\lambda} \gamma_5 \sigma_3 \lambda=\bar{\chi} \gamma_5 \chi-\bar{\eta}\gamma_5 \eta\,,\\
  d(0^{--})=&\bar{\lambda} \sigma_1 \lambda=\bar{\chi} \eta\,,\quad d(0^{-+})=\bar{\lambda} \sigma_3 \lambda=\bar{\chi} \chi-\bar{\eta}\eta\,,\\
  f(0^{++})=&\bar{\lambda}\lambda=\bar{\chi} \chi+\bar{\eta}\eta\,,\quad \eta(0^{-+})=\bar{\lambda}\gamma_5 \lambda=\bar{\chi} \gamma_5 \chi+\bar{\eta} \gamma_5 \eta.
 \end{aligned}
\end{equation}
Table \ref{tab:bilinearInvariants} shows their behaviour under the chiral transformations together with the corresponding Goldstone bosons.
\begin{table}
\begin{center}
\begin{tabular}{|c|c|c|c|c|c|c|}
\hline Operator & Parameter & $O_\mathsf{A,1}$ & $O_\mathsf{A,3}$ & $O_\mathsf{V,2}$ & Goldstone bosons & Massive state\\
\hline
\hline $\bar{\lambda}\lambda$ & $m$ & \uchk & \uchk & \chk & $d(0^{++})$ , $d(0^{+-})$ & $f(0^{++})$\\
\hline $\bar{\lambda}\gamma_5 \sigma_1\lambda$ & $\tilde{J_1}$ & \uchk & \chk & \uchk & $d(0^{++})$ , $f(0^{++})$ & $d(0^{+-})$\\
\hline $\bar{\lambda}\gamma_5 \sigma_3\lambda$ & $\tilde{J_2}$ & \chk & \uchk & \uchk & $d(0^{+-})$ , $f(0^{++})$ & $d(0^{++})$\\
\hline
\hline $\bar{\lambda} \gamma_5 \lambda$ & $m_5$ & \uchk & \uchk & \chk & $d(0^{-+})$ , $d(0^{--})$ & $\eta(0^{-+})$\\
\hline $\bar{\lambda}\sigma_1\lambda$ & $J_2$ & \uchk & \chk & \uchk & $d(0^{-+})$ , $\eta(0^{-+})$ & $d(0^{--})$\\
\hline $\bar{\lambda}\sigma_3\lambda$ & $J_1$ & \chk & \uchk & \uchk & $d(0^{--})$ , $\eta(0^{-+})$ & $d(0^{-+})$\\
\hline
\hline $\bar{\lambda}\gamma_0 \sigma_2 \lambda$ & $\mu$ & \uchk & \uchk & \chk &  - & -\\
\hline
\end{tabular}
\end{center}
\caption{The table shows the transformation behaviour of bilinears under the chiral transformations, \chk\, means invariant, \uchk\, not invariant. In the last two columns the corresponding Goldstone bosons and the massive states are shown.}
\label{tab:bilinearInvariants}
\end{table}
Under the chiral $SU(2)$ these biliners decompose as
\begin{equation}
 \bar{2} \otimes 2=3 \oplus 1,
\end{equation}
and we can identify a positive and negative parity triplet,
\begin{equation}
\begin{pmatrix}f(0^{++}), & d(0^{++}), & d(0^{+-})\end{pmatrix}\quad \text{and} \quad \begin{pmatrix}\eta(0^{-+}), & d(0^{-+}), & d(0^{--})\end{pmatrix}.
\end{equation}
Since the negative parity multiplet obtains a contribution to its mass from the chiral anomaly we will set the corresponding sources $J_1=J_2=m_5=0$.  
Under a general (infinitesimal) chiral transformation $\delta \lambda=\ii \left(\alpha \gamma_5 \sigma_1+\beta \gamma_5 \sigma_2+\gamma \sigma_2\right)\lambda$ the Lagrangian density transforms as
\begin{equation}
 \delta\mathcal{L}=2 \ii \bar{\lambda}\left(\gamma_5 \sigma_1 (-\alpha m+\gamma\tilde{J}_2)+\gamma_5 \sigma_3 (-\beta m-\gamma\tilde{J}_1)+\ii( \alpha \tilde{J}_1+\beta \tilde{J}_2)+\ii \mu \gamma_0 \gamma_5 (\beta \sigma_1-\alpha \sigma_2)\right)\lambda\, ,
 \end{equation}
 and we obtain the following system of equations for the invariance of the Lagrangian
 \begin{equation}
   \mu \alpha=0\,, \quad \mu \beta=0\,,\quad \alpha m=\gamma \tilde{J}_2\,, \quad \beta m=-\gamma \tilde{J}_1\,, \quad \alpha \tilde{J_1}=-\beta \tilde{J}_2.
 \end{equation}
 Possible solutions are shown in Table \ref{tab:solutions}.
 \begin{table}
 \begin{center}
\begin{tabular}{|c||c|c||c|c|c|}
\hline $\mu$ & $m$ & $\tilde{J}_{1,2}$ & Solution & Symmetry & Generator\\
\hline
\hline \chk &  \chk & \chk & & $SU(2)$ & $T=\{\gamma_5 \sigma_1,\gamma_5 \sigma_3,\sigma_2\}$ \\
\hline \chk &  \uchk & \chk & $\alpha=\beta=0$ & $U(1)$ & $T=\sigma_2$ \\
\hline \chk &  \chk & \uchk & $\gamma=0\,,\alpha \tilde{J_1}=-\beta \tilde{J}_2$ & $U(1)$ & $T=\gamma_5 (\tilde{J}_2 \sigma_1-\tilde{J}_1\sigma_3)$ \\
\hline \chk &  \uchk & \uchk & $\alpha m=\gamma \tilde{J_2},\beta m=-\gamma \tilde{J}_1$ & $U(1)$ & $T=\gamma_5 (\tilde{J}_2\sigma_1-\tilde{J}_1\, \sigma_3)+m \sigma_2$ \\
\hline
\hline \uchk &  \chk & \chk & $\alpha=\beta=0$ & $U(1)$ & $T=\sigma_2$ \\
\hline \uchk &  \uchk & \chk & $\alpha=\beta=0$ & $U(1)$ & $T=\sigma_2$ \\
\hline \uchk &  \chk & \uchk & $\alpha=\beta=\gamma=0$ & - & - \\
\hline \uchk &  \uchk & \uchk & $\alpha=\beta=\gamma=0$ & - & - \\
\hline
\end{tabular}
\end{center}
\caption{Chiral symmetry of $G_2$-QCD in the presence of diquark sources and chemical potential,\chk\, means $=0$, \uchk\ means $\neq 0$.}
\label{tab:solutions}
\end{table}
The Goldstone bosons are then linear combinations in the corresponding multiplet. An illustration of the chiral symmetry breaking is shown in Figure~\ref{fig:chiralSymmDiquark}.
At vanishing chemical potential, any linear combination of mass $m$ and diquark sources $\tilde{J}_1$ and $\tilde{J}_2$ breaks the chiral symmetry down to a $U(1)$ subgroup and baryon number is conserved. The direction (generator) for the invariant subgroup $U(1)$ of
the $SU(2)$ symmetry is shown in Table~\ref{tab:solutions}. Since the mass $m$ and the chemical potential $\mu$ break the $SU(2)$ in the same direction, baryon number is explicilty broken at finite chemical potential by any
linear combination of diquark sources. Thus in our simulations at zero density we expect no qualitative difference between a fermion mass term and a diquark source term.

\section{Results at finite temperature}
\begin{figure}[tb]

\begin{center}

\vspace*{.2cm}

\scalebox{0.61}{\input{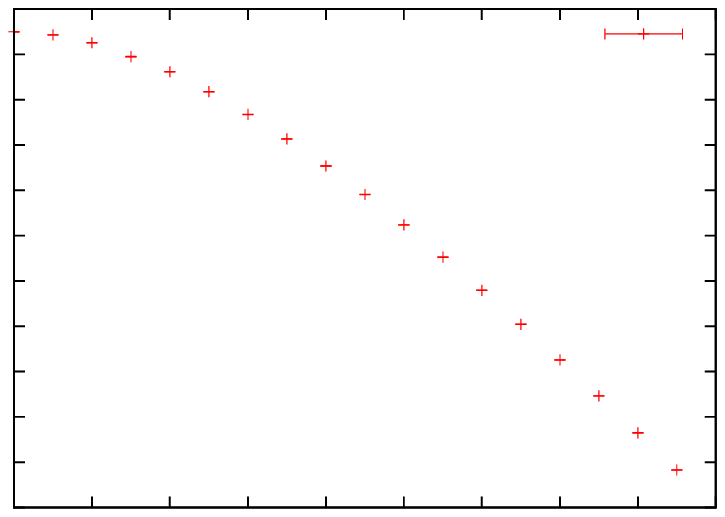}}\hskip02mm
\scalebox{0.61}{\input{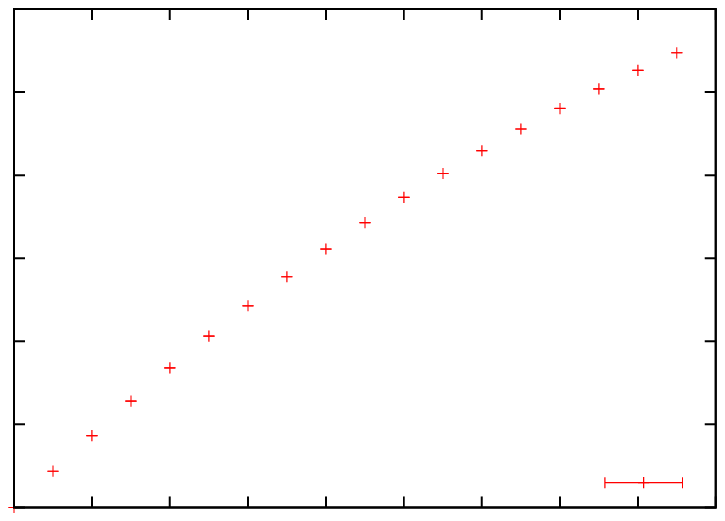}}\hskip02mm
\scalebox{0.61}{\input{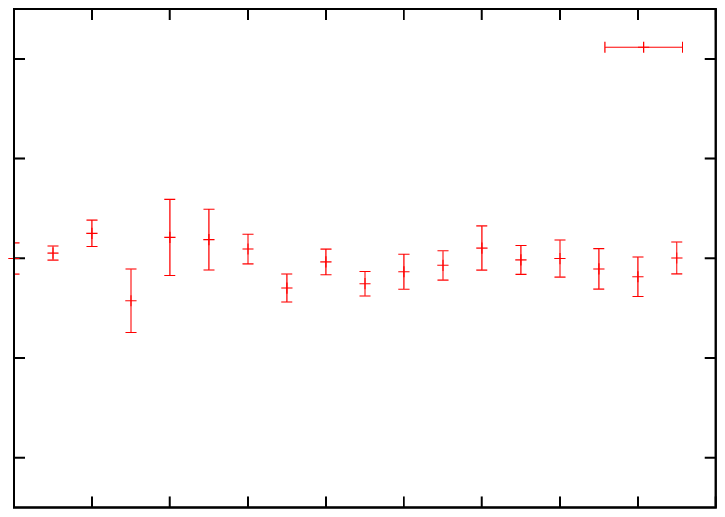}}
\end{center}
\caption{Chiral condensate, diquark condensate and quark number density as a function of the diquark source coupling $\tilde{J}_1$ in the vacuum at $T=\mu=0$.}
\label{fig:Majorana_vacuum}
\end{figure}

\begin{figure}[htb]
\begin{center}
\scalebox{0.60}{\input{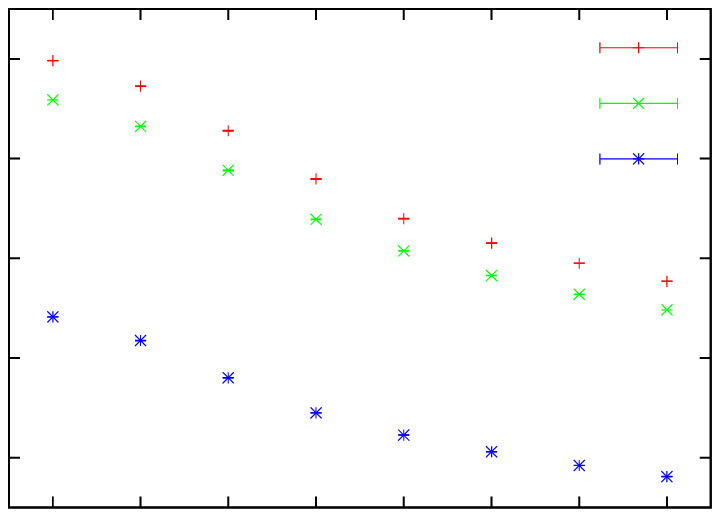}}\hskip03mm
\scalebox{0.60}{\input{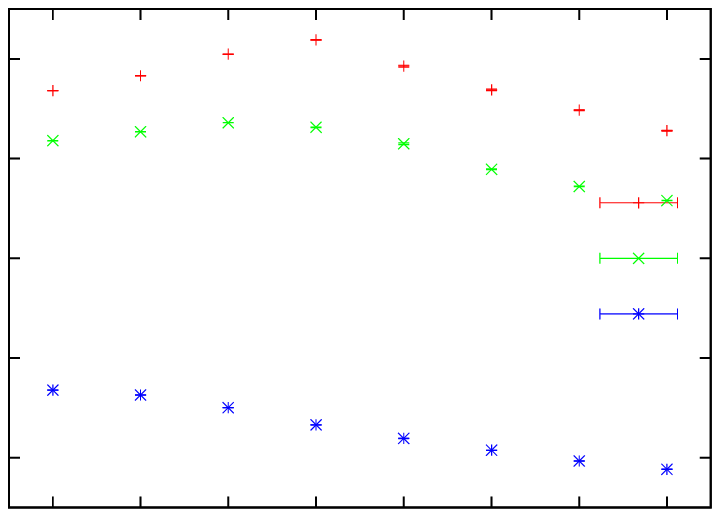}}\hskip03mm
\scalebox{0.60}{\input{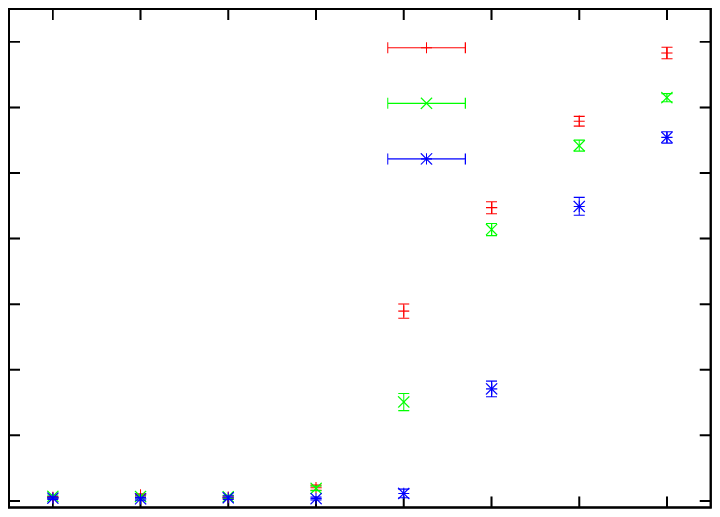}}
\end{center}
\caption{Chiral condensate (left), diquark condensate (in units of $\tilde J_1$, middle) and Polyakov loop (right) as a function of the inverse gauge coupling $\beta$ for three different values of the diquark source $\tilde J_1$.}
\label{fig:Majorana_FiniteT}
\end{figure}

\begin{figure}[b]
\begin{center}
\scalebox{0.85}{\input{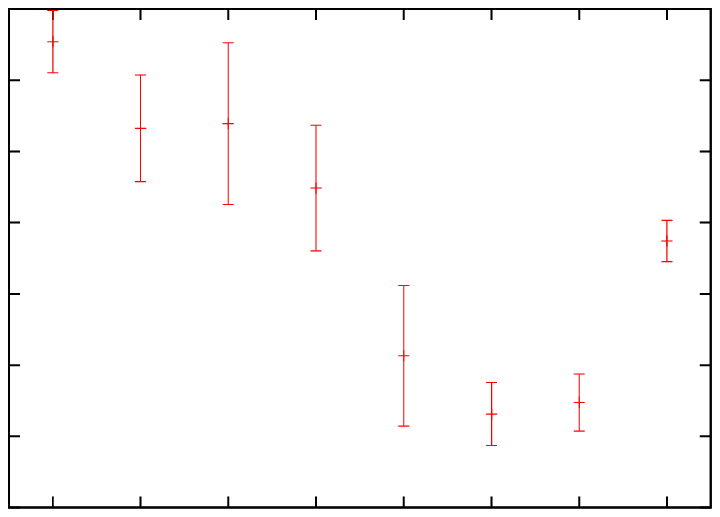}}\hskip10mm
\scalebox{0.85}{\input{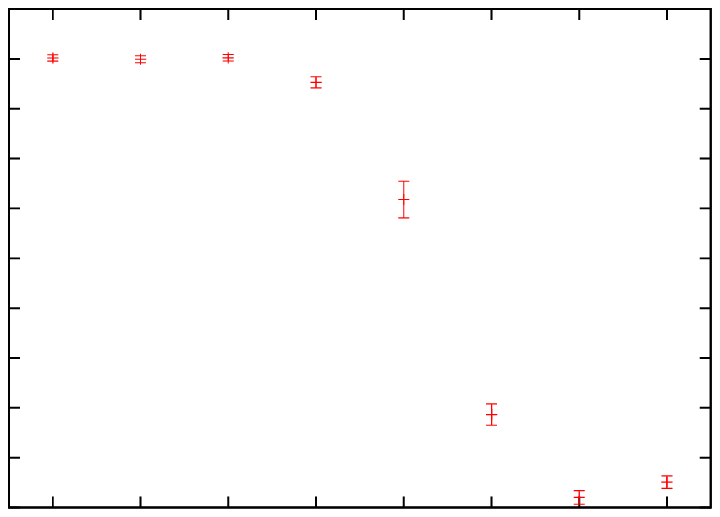}}
\end{center}
\caption{Normalized chiral (left) and diquark (right) condensates over the inverse gauge coupling $\beta$.}
\label{fig:Majorana_FiniteT_ren}
\end{figure}

 \noindent
 Our first aim is to investigate spontaneous chiral symmetry breaking at finite temperature for $\mu=0$. In our simulations with Dirac fermions the 
corresponding signal in the chiral condensate remained rather inconclusive when measured across the deconfinement transition as observed in the Polyakov loop \cite{Maas:2012wr}. The reason for the weak transition in the chiral condensate was probably the explicit breaking due to the Wilson mass.
 The first simulations with Majorana fermions at zero temperature have been performed on a $8^3 \times 16$ lattice with $\beta=0.96$ and $\kappa=0.151$. Compared to the ensembles discussed for Dirac fermions before, this corresponds to very heavy quarks (diquarks).
 For vanishing chemical potential the Pfaffian is again positive and simulations can be done.
 In Figure \ref{fig:Majorana_vacuum} the chiral condensate, the diquark condensate and the quark number density are shown as a function of the diquark source $\tilde{J}_1$.
As expected, with increasing diquark source (and fixed mass $m$), the chiral condensate rotates into a diquark condensate. Furthermore, the quark number density vanishes, indicating an unbroken $U(1)$ subgroup such that baryon number is still conserved. Our simulations at finite temperature have been performed on a $12^3 \times 6$ lattice with $\beta=0.96$ and $\kappa=0.156$. In Figure~\ref{fig:Majorana_FiniteT}, the chiral condensate, diquark condensate and the Polyakov loop
are shown as a function of the inverse gauge coupling $\beta$ for three different values of the diquark source. 
With increasing diquark source the deconfinement transition visible in the Polyakov loop is shifted to larger values of $\beta$.
This is the same qualitative behaviour as observed when increasing the fermion mass. In the chiral condensate and the diquark condensate only a very weak transition is observed which becomes more pronounced for smaller diquark masses.
For a particular diquark source of $\tilde{J}_1=0.20$ we have normalized the condensates by their vacuum contribution. The result is shown in Figure~\ref{fig:Majorana_FiniteT_ren}.
The transition in the chiral condensate is very weak while it is by a factor of $\sim 40$ stronger in the diquark condensate. We have so far ignored the necessity for additive renormalization in the chiral condensate, however. Also the error bars are much smaller for the diquark condensate as compared to the chiral condensate.
Compared to the situation without diquark sources we observe a chiral improvement like for instance in twisted mass QCD where a similar term is added to the action in order to improve the chiral properties.
The transition temperature agrees with the one obtained from the Polyakov loop. A future task in our simulations with Majorana fermions at finite temperature will be to repeat the spectroscopy for smaller fermion masses and various values for the diquark source term.

\section{Results at finite density}
\noindent
At finite density we have no argument that the Pfaffian should remain positive if we add a diquark source, but we do no expect a severe sign problem for small baryon densities.
Our first simulations here were done on an $8^3 \times 16$ lattice with $\beta=0.96$ and $\kappa=0.156$.
\begin{figure}[t]
\begin{center}
\scalebox{0.62}{\input{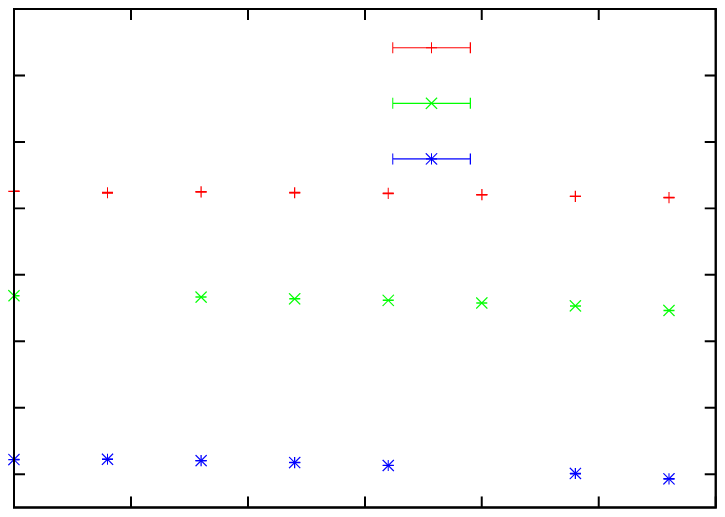}}\hskip01mm
\scalebox{0.62}{\input{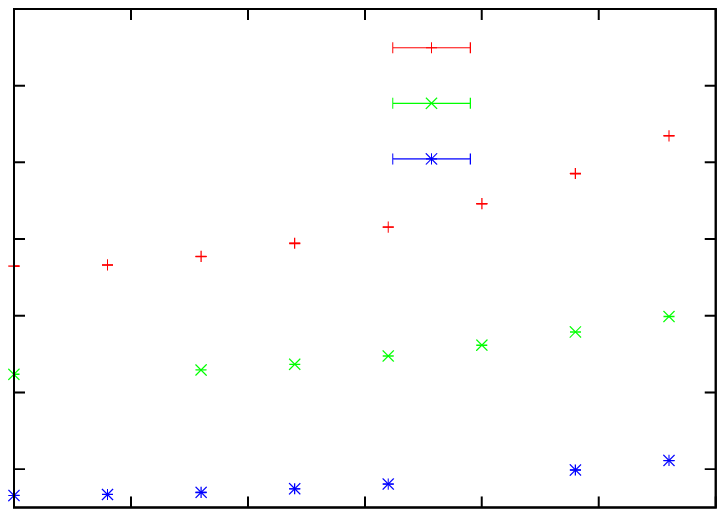}}\hskip01mm
\scalebox{0.62}{\input{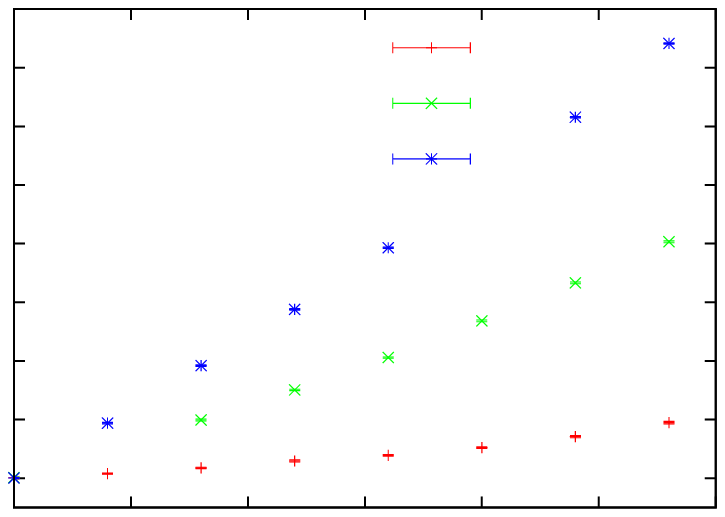}}
\end{center}
\caption{Chiral condensate (left), diquark condensate (in units of $\tilde J_1$, middle) and quark number density (right) over chemical potential at $T=0$.} 
\label{fig:Majorana_FiniteMu}
\end{figure}
Figure~\ref{fig:Majorana_FiniteMu} shows the chiral condensate, the diquark condensate and the quark number density as a function of small values of the chemical potential (below the Silver Blaze transition) and different values for the diquark source. Here we observe that baryon number (and chiral symmetry) is explicitly broken at non-zero $\mu$ such that the quark number density is always non-zero for $\mu\not=0$. We intend to measure the diquark spectrum as a function of chemical potential in order to further test our assumptions about chiral symmetry breaking at finite density in future.

\section{Conclusions}\label{sconclusions}
\noindent
In this contribution to Lattice 2014 we have reviewed our present results for the mass spectrum of $G_2$-QCD and its finite density phases at zero temperature.
For small quark masses we observe a splitting of the spectrum into a pseudo-Goldstone sector, an intermediate bosonic baryon sector, and a nucleonic sector. Although the spectrum of $G_2$-QCD is much richer than that of ordinary QCD, the results relevant for hadronic physics appear to be quite similar to QCD.
In the phase diagram at zero temperature one observes several transitions in the baryon or quark number density that correlate well with the different mass scales in the baryon spectrum. The transition with the smallest value of the chemical potential coincides with the mass of the lightest state carrying baryon number, consistent with the Silver Blaze property which is realized in 
 $G_2$-QCD as in two-color QCD or QCD at finite isospin chemical potential. The zero-density ground state remains unchanged until the quark chemical potential reaches half the mass of the pseudo-Goldstone boson which is the scalar diquark as in two-color QCD, a baryon with quark number two.
 
 We also find evidence that a phase dominated by fermionic baryons exists for quark chemical potentials above about 300-600 MeV.
In between we found good evidence of a first-order phase transition from a phase that is dominated by diquark matter to a phase that is probably dominated by ordinary baryonic matter.
This transition occurs somewhat below the scale given, in units of the quark number, by the nucleon mass as obtained from our vacuum spectroscopy. This might indicate a rather large binding energy per nucleon or that, for densities above the Silver Blaze transition, the spectrum of the theory depends stronger on the chemical potential than one might expect.

In order to investigate the chiral properties of the theory in more detail, we have introduced diquark sources, in particular, to disentangle the pseudo-Goldstone boson spectrum. This led us to simulate Majorana fermions and allowed  
 us to explicitly demonstrate the vacuum realignment at zero temperature and chemical potential when the strength of the corrsponding diquark source term increases relative to the fermion mass. Moreover, it turned out, that the signal for the finite-temperature chiral transition at vanishing net-baryon density is considerably stronger in the diquark condensate than it is in the standard chiral condensate. Possible explanations are the explicit breaking of chiral symmetry by the Wilson mass that competes with the spontaneous breaking of the chiral condensate and the necessity of additive renormalizations. This behaviour is reminiscent of simulations with twisted-mass fermions in QCD where the chiral properties of the theory are improved by analogous terms in the fermion action.

\begin{acknowledgments}
\noindent 
Discussions and collaboration with Axel Maas and Andreas Wipf are gratefully acknowledged. 
This work was supported by the Helmholtz International Center for FAIR within the LOEWE initiative of the State of Hesse.
Simulations were performed on the LOEWE-CSC at the University of Frankfurt.
\end{acknowledgments}


\begin{thebibliography}{99}

\providecommand{\eprint}[1]{ [\href{http://arxiv.org/abs/#1}{arXiv:#1}]}

\bibitem{Holland:2003jy}
  K.~Holland, P.~Minkowski, M.~Pepe and U.J.~Wiese,
  \newblock Nucl.\ Phys.\ B {\bf 668} (2003) 207 [hep-lat/0302023].

\bibitem{Maas:2012wr}
  A.~Maas, L.~von Smekal, B.H.~Wellegehausen and A.~Wipf,
  \newblock Phys.\ Rev.\ D {\bf 86} (2012) 111901(R)     
  [arXiv:1203.5653 [hep-lat]].

\bibitem{Pepe:2006er}
  M.~Pepe and U.-J.~Wiese,
  Nucl.\ Phys.\ B {\bf 768} (2007) 21 [hep-lat/0610076].

\bibitem{Greensite:2006sm}
 J.~Greensite, K.~Langfeld, S.~Olejnik, H.~Reinhardt and T.~Tok,
 Phys.\ Rev.\ D {\bf 75} (2007) 034501 [hep-lat/0609050].

\bibitem{Danzer:2008bk}
  J.~Danzer, C.~Gattringer and A.~Maas,
  JHEP {\bf 0901} (2009) 024 [arXiv:0810.3973 [hep-lat]].

\bibitem{Wellegehausen:2010ai}
  B.H.~Wellegehausen, A.~Wipf and C.~Wozar,
  Phys.\ Rev.\ D {\bf 83} (2011) 016001 [arXiv:1006.2305 [hep-lat]].

\bibitem{Ilgenfritz:2012wg}
  E.M.~Ilgenfritz and A.~Maas,
  Phys.\ Rev.\ D {\bf 86} (2012) 114508
  [arXiv:1210.5963 [hep-lat]].

\bibitem{Lacroix:2012pt}
  G.~Lacroix, C.~Semay, D.~Cabrera and F.~Buisseret,
  Phys.\ Rev.\ D {\bf 87} (2013) 054025 [arXiv:1210.1716 [hep-ph]].
 
\bibitem{Maas:2012ts}
  A.~Maas and B.H.~Wellegehausen,
  \newblock PoS LATTICE2012 (2012) 080 [arXiv:1210.7950 [hep-lat]].

\bibitem{Wellegehausen:2013cya}
  A.~Maas, L.~von Smekal, B.H.~Wellegehausen and A.~Wipf, 
  \newblock Phys.\ Rev.\ D {\bf 89} (2014) 056007
  [arXiv:1312.5579 [hep-lat]].

\bibitem{Kogut:2000ek}
  J.B.~Kogut, M.A.~Stephanov et al.,
 \newblock Nucl.\ Phys.\ B {\bf 582} (2000) 477 [hep-ph/0001171].
 
 \bibitem{Hands:2000ei}
  S.~Hands, I.~Montvay, S.~Morrison, M.~Oevers and L.~Scorzato,
  \newblock Eur.\ Phys.\ J. C {\bf 17} (2000) 285 [hep-lat/0006018].
  



 


\end{thebibliography}
\end{document}